\documentclass[aps,prd,12pt,preprint,nofootinbib,superscriptaddress,titlepage,tightenlines,floatfix]{revtex4}

\usepackage{graphicx}
\usepackage{dcolumn}
\usepackage{bm}

\newcommand{\BABAYAGA}{{\sc Babayaga}}
\newcommand{\BABAR}{{\sc BaBar}}
\newcommand{\Is}{I(s)}
\newcommand{\Iz}{I_0}

\newcommand{\diel}{e^+e^-}
\newcommand{\dimu}{\mu^+\mu^-}
\newcommand{\dipion}{\pi^+\pi^-}

\newcommand{\jpsi}{{J/\psi}}

\newcommand{\psiprime}{\psi(2S)}
\newcommand{\gamee}{\Gamma_{ee}}
\newcommand{\gamtot}{\Gamma_{\rm tot}}

\newcommand{\sprime}{s^{\,\prime}}
\newcommand{\bprime}{B^{\,\prime}}
\newcommand{\ie}{{\sl i.e.,}}

\newcommand{\br}{{\cal B}}
\newcommand{\brmm}{{\cal B}_{\mu\mu}}
\newcommand{\bree}{{\cal B}_{ee}}
\newcommand{\brll}{{\cal B}_{\ell\ell}}
\newcommand{\nj}{N_\jpsi}
\newcommand{\nb}{N_{\rm bgd}}
\newcommand{\mj}{M_\jpsi}
\newcommand{\dndm}{N_{\rm QED}} 
\newcommand{\dsdm}{\delta_{\rm QED}} 
\newcommand{\lumi}{{\cal L}}
\newcommand{\ezero}{\epsilon_{\rm QED}}
\newcommand{\azero}{a_{0}}
\newcommand{\szero}{s_{0}}
\newcommand{\sigzero}{\sigma_{0}}

\begin{document}

\preprint{CLNS 05/1945}
\preprint{CLEO 05-31}

\title{Measurement of $\Gamma_{ee}(J/\psi)$, $\Gamma_{\rm tot}(J/\psi)$, and $\Gamma_{ee}[\psi(2S)]/\Gamma_{ee}(J/\psi)$}

\author{G.~S.~Adams}
\author{M.~Anderson}
\author{J.~P.~Cummings}
\author{I.~Danko}
\author{J.~Napolitano}
\affiliation{Rensselaer Polytechnic Institute, Troy, New York 12180}
\author{Q.~He}
\author{J.~Insler}
\author{H.~Muramatsu}
\author{C.~S.~Park}
\author{E.~H.~Thorndike}
\affiliation{University of Rochester, Rochester, New York 14627}
% ...
\collaboration{CLEO Collaboration} 
\noaffiliation

\date{December 16, 2005}

\begin{abstract} 
 Using data acquired with the CLEO detector at the CESR $\diel$ collider
at $\sqrt{s}=3.773$~GeV, we 
measure the cross section for 
the radiative return process $\diel\to\gamma \jpsi$, 
$\jpsi\to\dimu$, resulting in
$\br(\jpsi\to\dimu)\times\gamee(\jpsi)=0.3384\pm0.0058\pm0.0071$~keV,
$\gamee(\jpsi)=5.68\pm 0.11 \pm 0.13$~keV, and
$\gamtot(\jpsi)=95.5\pm 2.4\pm 2.4$~keV,
in which the errors are statistical and systematic, respectively.
We also determine the ratio
$\gamee[\psiprime]/\gamee(\jpsi)=0.45\pm0.01\pm0.02$.
\end{abstract}

\pacs{13.20.Gd,14.40.Gx}
\maketitle

The full and dileptonic widths of a hadronic resonance, 
$\gamtot$ and $\gamee$, describe
fundamental properties of the strong potential~\cite{QWG}. The value of 
$\gamee$ for a particular resonance is, in principle,
predictable within QCD, although the strong interaction
effects in the quark-antiquark pair annihilation make
calculations challenging. Heavy quarkonia offer the best
testing ground for lattice-based (LQCD) techniques~\cite{LQCD}, 
and a fortuitous convergence in precision to the few percent level 
is occuring on both the theoretical and experimental fronts. 
In 2003, \BABAR\ measured 
$\brmm\times\gamee$~\cite{babar}, where $\brmm\equiv\br(\jpsi\to\dimu)$,
using the novel technique of counting radiative returns 
from $\diel$ collisions at $\sqrt{s}=10.58$~GeV.
This allowed the world average~\cite{PDG04} uncertainty on
$\gamee(\jpsi)$ and $\gamtot(\jpsi)$
to be reduced by nearly a factor of two
when combined with a BES~\cite{BESJLL} 
determination of the $\jpsi$ dileptonic branching fraction $\brll$,
which has a relative 1.7\% uncertainty. At the same time,
progress on predicting and measuring dielectronic widths of
bottomonium~\cite{LQCD,GAMUPS} is occuring, providing further checks of
LQCD computations. For the $\Upsilon$ system, the LQCD predictions
for the ratios $\gamee[\Upsilon(nS)]/\gamee[\Upsilon(1S)]$
are expected to be more accurate than those of the absolute individual
widths.

  In this Article we describe a measurement of the $\jpsi$
full and dielectronic widths with CLEO-c data
from $\diel$ collisions near the peak of the $\psi(3770)$ resonance. 
The method is similar to
that used earlier by \BABAR~\cite{babar} and 
in CLEO's recent measurement~\cite{xj3770prl} 
of $\gamee[\psiprime]$: we select $\dimu(\gamma)$
events, each with a dimuon mass in the general region of the $\jpsi$,
and count the excess over non-resonant QED production, $\diel\to\gamma\dimu$.
(The $\diel$ final state is not used due to the large
$t$-channel contribution, which limits the attaintable statistical
precision relative to $\dimu$).
The $\jpsi$ component will peak at $M(\dimu)$=$\mj$ with a
mass resolution dominated by detector effects.
The cross section for the excess is proportional to 
$\brmm\times\gamee(\jpsi)$. Assuming lepton universality,
we can then divide by CLEO's own $\brll$~\cite{dilepprd},
with a relative accuracy of 1.18\%, once to obtain
$\gamee(\jpsi)$ and once more for $\gamtot(\jpsi)$.

We use $\diel$~collision data 
collected with the CLEO detector~\cite{CLEOdetector}
acquired at a center-of-mass energy $\sqrt{\szero}$=3.773~GeV 
at the Cornell Electron Storage Ring (CESR)~\cite{cesr}. 
The CLEO detector features a solid angle coverage of $93\%$ for
charged and neutral particles. 
The charged particle tracking system operates in a 1.0~T~magnetic field
along the beam axis and achieves a momentum resolution of
$\sim$0.6\% at momenta of $1$~GeV/$c$. 
The integrated luminosity ($\cal{L}$) was measured 
using $\diel$, $\gamma\gamma$, and $\dimu$
events~\cite{LUMINS} and normalized with 
a Monte Carlo (MC) simulation based on the \BABAYAGA~\cite{babayaga} 
generator combined with GEANT-based~\cite{GEANT} detector modeling. 
Results from the three final states are consistent and
together yield ${\cal L}=280.7\pm2.8$~pb$^{-1}$.

 The differential
cross section for ${\diel\to \gamma \jpsi}\to \gamma \dimu$
can be expressed~\cite{KF,radcor,vectorisr} in terms of the $\diel$
invariant mass $s$, the
dimuon mass-squared $\sprime$, and the variable $x\equiv (1-\sprime/s)$ as
\begin{equation}
{{d\sigma}\over{dx}}(s,x)= W(s,x)\times b(\sprime)\times 
\gamee\times\brmm\ \ ,
\end{equation}
\noindent where $W(s,x)$ is the initial state radiation (ISR) 
$\gamma$-emission probability, 
$b(\sprime)$ is the relativistic Breit-Wigner function, 
and  $\gamee$ is the $\jpsi$ 
$\diel$ partial width (including vacuum polarization effects).
The ISR kernel,
to lowest order in the fine structure constant $\alpha$, is
\begin{equation}
W(s,x)\equiv {{2\alpha}\over{\pi
x}}\left(\ln{s\over{m_e^2}}-1\right)\left(1-x+{{x^2}\over2}\right)\, ,
\end{equation}
\noindent in which $m_e$ is the electron mass. The Breit-Wigner function is 
$b(\sprime)\equiv B(\sprime)/\brmm\gamee$, 
\begin{equation}
B(\sprime)\equiv { {12\pi \brmm\gamee \gamtot }\over{
(\sprime-M^2)^2+ M^2\gamtot^2 } }\ ,
\end{equation}
\noindent where $\gamtot$ is the full width
and $M$ the $\jpsi$ mass. 

  The cross section $\sigzero\equiv\sigma(\szero)$ for
$\diel\to \gamma \jpsi\to \gamma \dimu$ over a specified
dimuon mass range
can be obtained from Eq.~(1) and measured:
\begin{equation}
\sigzero = {{ \nj-\nb}\over{\epsilon\times {\cal L}}} =
\gamee\times\brmm\times \Iz \ ,
\end{equation}
\noindent in which $\nj$ is the number of signal
events counted, $\nb$ is the estimated background, $\epsilon$
is the detection efficiency obtained from Monte Carlo (MC)
simulation, $\Iz\equiv I(\szero)$, and the integral
\begin{equation}
\Is\equiv \int W(s,x)\ 
 {{b(\sprime)}}\  dx
\end{equation}
\noindent  is effectively insensitive to the value of $\gamtot$. 
Hence a measurement of $\sigzero$ 
can be combined with 
$\brmm$ measurements~\cite{dilepprd} to yield 
$\gamee(\jpsi)$. For these equations to work,
the number of events $N$ and the integral above must
both be determined with the same limits on $x$,
which means the same limits on muon pair mass.

  The above treatment ignores interference effects with
the QED $\dimu$ production,
which modify the dimuon mass lineshape asymmetrically around the peak; 
these effects are included at lowest order by replacing $B(\sprime)$ with
\begin{equation}
\bprime(\sprime)\equiv { {4\pi \alpha^2 }\over{3\sprime}}
\left( \ \left| 1-{{Q\,M\gamtot}\over{M^2-\sprime-iM\gamtot}} \right|^2 - 1\right),
\end{equation}
\noindent where $Q\equiv 3\sqrt{\bree\brmm}/\alpha$.

Evaluation of Eq.~(6) shows that interference
is constructive above the peak and symmetrically 
destructive below, inducing a change in the
Breit-Wigner cross section at $\sqrt{s}=\mj\pm\gamtot/2$, 
for example, of about $\pm8$\%. At $\pm300$~MeV
from the peak, the interference term induces an effect equal to $\pm0.8\%$ 
of the nonresonant QED $\dimu$ cross section.

The integral $\Is$ is performed numerically.
The result for our $\sqrt{s}$=3.773~GeV dataset
and $x$=0.139-0.488 (\ie~$M(\dimu)$=2.8-3.4~GeV) 
is $\Iz$=188.8$\pm$1.3~pb/keV
using $W(s,x)$ from Eq.~(2) and $\Iz$=185.8$\pm$1.3~pb/keV
when including the radiative corrections in Eq.~(28) of Ref.~\cite{KF}. 
Both values for $\Iz$ include a net relative increase of $\sim$3\% 
due to the 
interference term (using Eq.~(6) instead of Eq.~(3)),
which occurs because the ISR kernel's $1/x$ term weights
higher masses (constructive interference) more than lower masses
(destructive interference). The quoted uncertainties on $\Iz$ are based only
on the statistics of the numerical integrations.

   The event selection procedure is straightforward. The 
two highest-momentum tracks are required to have opposite charge,
individually satisfy either $|\cos\theta|$$<$0.83 or  0.85$<$$|\cos\theta|$$<$0.93
so as to avoid the barrel-to-endcap calorimeter transition region, 
and together have an invariant mass in the range 2.8-3.4~GeV.
Bremsstrahlung photons, defined as calorimeter showers
found within a 100~mrad cone about the initial charged track direction,
are added to the corresponding Lorentz vector for
the $M(\dimu)$ computation for each event.
Muon pairs are loosely selected, and electrons effectively vetoed, 
by requiring the two tracks to satisfy muon-like requirements on the 
matched energy-to-momentum ratio $E/p$: 
the larger of the two $E/p$ values must be
$<$0.5 and the smaller $<$0.25; electrons typically
have $E/p\simeq$1 and consistently satisfy $E/p>$0.5.
Cosmic rays are suppressed by
requiring the pair of tracks together to point to within 2~mm of zero
in the plane perpendicular to the beams and 
within 40~mm of zero along the beam direction.
Cosmic rays are further suppressed 
by requiring the candidate $\jpsi$ to have momentum
$p_{\mu\mu}=0.1$-1.5~GeV/$c$, a restriction which has 
no effect upon signal efficiency.

\begin{figure}[thp]
\includegraphics*[width=6.5in]{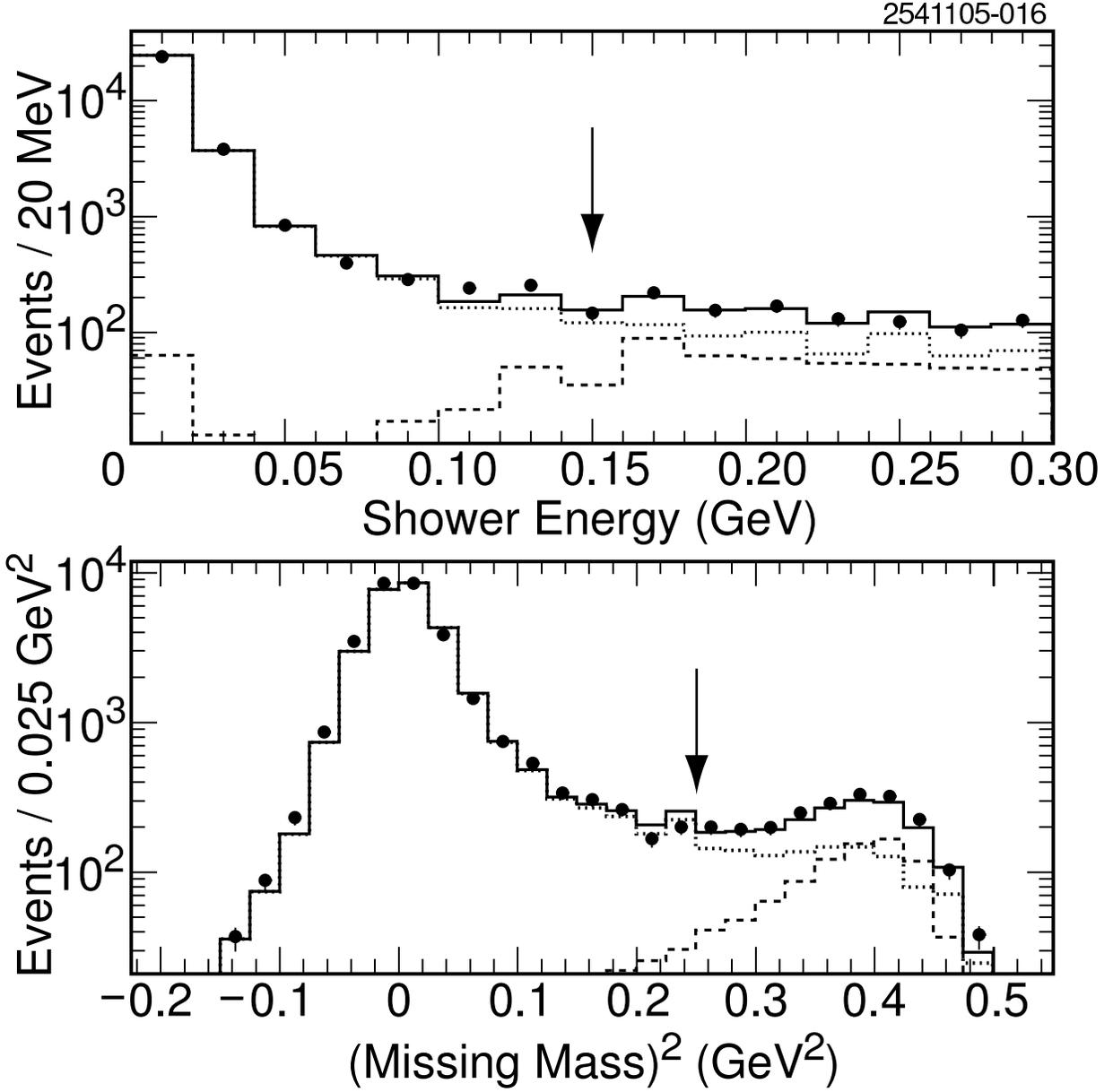}
\caption{Distributions, for $M_{\mu\mu}$=3.05-3.15~GeV, of the largest calorimeter 
shower energy unaffiliated with
a charged track (top)
and missing-mass-squared (bottom) 
for the data (filled circles),
in the signal $\gamma \jpsi$ MC (dotted line histogram),
$\gamma\psiprime\to\gamma X \jpsi$ MC (dashed), and their sum (solid). 
Arrows show the nominal upper limits of values accepted by 
the event selection.\label{fig:cut} }
\end{figure}

   The dominant backgrounds to a $\gamma\jpsi$ signal are radiative
returns to $\psiprime$ with subsequent decays $\psiprime\to X\jpsi$. 
Such events will have a true $\jpsi\to\dimu$,
but they will also tend to have extra tracks
or showers, as well as a significant mass recoiling against
the muon pair. Three requirements are imposed to suppress these
events: one on extra charged tracks, a second on extra calorimeter 
showers, 
and a third on missing mass.
The number of tracks satisfying
loose quality criteria is required to be exactly two.
The missing mass,
$| p_{\rm cm} - ( p_{\mu^+} + p_{\mu^-} ) | $,
where $p_{\rm cm}$ is the initial state
center-of-mass Lorentz vector, and $p_{\mu^+}$ and $p_{\mu^-}$
are the two muon Lorentz vectors, is required 
to be less than 500~MeV; this value is set by
the need to reject both charged and neutral $\pi\pi \jpsi$ events. 
We search for the most energetic shower unassociated
with a charged track
that is not within a 100~mrad cone of either the initial momentum
direction of either track 
or of the opposite of the net muon pair momentum direction, 
and demand it to have energy below 150~MeV. 
Figure~\ref{fig:cut} shows the two variables
sensitive to backgrounds from radiative returns to $\psiprime$
for signal and $\gamma\psiprime\to \gamma X\jpsi$, $\jpsi\to\dimu$ MC, 
demonstrating that the restrictions on extra showers
and missing mass separate the signal
from these backgrounds.
 
The {\tt EvtGen} event generator~\cite{EVTGEN}, which includes
final state radiation (FSR)~\cite{PHOTOS}, and a 
GEANT-based~\cite{GEANT} detector simulation
are used to study the radiative return (to $\jpsi$ and $\psiprime$) 
processes with exactly one ISR photon.
Events are generated
with the polar angle distribution from Ref.~\cite{vectorisr}, 
and account for ISR according to Eqs.~(1-3).
Non-resonant events of the type $\gamma(\gamma...)\dimu$ are generated from
the \BABAYAGA~\cite{babayaga} package, which, unlike the
radiative return process of {\tt Evtgen}, includes the effects
of multiple ISR and/or FSR photon emission, and hence of higher orders
in $\alpha$. The \BABAYAGA\ code normally includes
the interference effects with $\jpsi$ decays, but
this feature was removed for the results shown here.

  In the absence of an MC generator package incorporating
multiple photon emission from the initial state and 
from $\jpsi$ decays in the radiative
return process, we calculate the efficiency of the
$\jpsi$ signal in three steps, assuming that ISR
and decay radiation are factorizable. For the first
step, in order to simulate production of
$\gamma(\gamma...)\jpsi\to\gamma(\gamma...)\dimu$, we use a subset of \BABAYAGA-generated
$\diel\to\gamma(\gamma...)\dimu$ events with FSR disabled
 and a muon pair mass (including photons emitted within
100~mrad of either muon's direction) restricted to within $\pm$10~MeV of
$\mj$. After detector simulation and reconstruction,
73.2\% of such events pass our selection criteria.
For the second step, we compare the {\tt EvtGen} efficiency 
with $\jpsi$ decay radiation from
PHOTOS~\cite{PHOTOS} to that without it, finding that
decay radiation reduces the efficiency by factor of 0.968.
The third step accounts for imperfections in modeling track-finding
and decay radiation, for which we correct the efficiency by
the factor 0.995~\cite{xjpsiprl}, arriving at $\epsilon$=70.5\%.

In order to probe both the background levels as well
as the modeling of the largest shower and missing mass restrictions,
we also perform the analysis with two alternate sets of
selection criteria, differing from nominal only in that for
the ``loose'' (``tight'') set, we require the highest energy
shower to have energy less than 200~MeV (100~MeV),
and that the missing mass be less than 550~MeV (450~MeV).
These variations result in a relative efficiency change of $+$1.6\% ($-$1.9\%).

  Table~\ref{tab:tableBgd} shows the expected number of
background events from $\gamma\psiprime\to \gamma X \jpsi$, $\jpsi\to\dimu$,
and from $\gamma \jpsi$, $\jpsi\to\dipion$. The total
number is 1.3\% of the signal. For
the alternate ``loose'' (``tight'') selection,
 the relative background prediction
is 5\% (0.5\%).
Other $\jpsi$ decay modes are found to contaminate
our sample at negligibly small levels.
No other processes will produce a peak at the $\jpsi$ mass.
Other backgrounds are assumed to be smooth in $M(\dimu)$ and fittable 
by a low-order polynomial.

  In order to avoid depending on near-perfect MC simulation of
the mass resolution, an alternate procedure is used to
generate an accurate expected shape of the dimuon mass spectrum.
We take a clean sample of $\jpsi\to\dimu$ decays
from $data$ in which there is no interference 
and convolve the measured mass resolution
with the expected effects from interference to obtain
the expected shapes. CLEO has already accumulated a 
large sample of essentially
background-free  radiative
return events from $\sqrt{s}$=3.773~GeV to $\psiprime$, 
$\psiprime\to\pi\pi \jpsi\to\pi\pi\dimu$~\cite{xj3770prl}, 
with almost the
same selection criteria as for this analysis. 
After rejecting events failing the 150~MeV unaffiliated shower veto, 
we take the mass distribution from these
11,305 events, summed over
both the charged and neutral dipion samples, and offset it by $\mj$ 
so as to be peaked at zero. The resulting distribution is taken
to represent the mass resolution function in the 2.8-3.4~GeV mass
region. In a toy MC, three different mass distributions are generated for
dimuons with ISR: from a $\jpsi$ decay alone (Eq.~(3)), from
non-resonant first-order QED, 
$\diel\to\gamma\dimu$ ($4\pi\alpha^2/3\sprime$), and for the
combination including interference ($4\pi\alpha^2/3\sprime +
\bprime(\sprime)$). 
For each ``event'' in each
distribution, the mass is smeared according to the mass resolution function
from $\gamma\psiprime$ data and recorded in a histogram. The final step is the subtraction of
the properly normalized QED-only nonresonant mass distribution from 
that for QED-plus-$\jpsi$-with-interference to obtain
the expected shape of the $\jpsi$ mass peak from radiative returns.
We designate this expected shape of the $\jpsi$ peak
including resolution and interference as $H(M_i)$, 
which has the property $\sum_i H(M_i)=1$, 
where $M_i$ is the center of the $i^{\rm th}$ mass bin.

  The following approach is taken for fitting the
smooth non-resonant
background: in seven fits from the widest window (2.8-3.4~GeV)
to the narrowest (3.06-3.14~GeV), the lowest order
polynomial is used that, in combination with the
signal shape from above, gives at least a 1.0\% confidence
level (CL) for the fit. In practice, this meant using a third
order polynomial for ranges wider than 2.9-3.3~GeV,
a linear background for ranges narrower than 3.03-3.17~GeV,
and a second order polynomial otherwise. We chose this strategy
to allow for statistical fluctuations in the signal resolution
function, to avoid introduction of an unphysical background shape,
and to maximize the orthogonality of the signal and background functions.
The fitting function is
\begin{equation}
f(M_i)=\nj H(M_i) + \Delta M\sum_{j=0}^3 a_j (M_i-\mj)^j ,
\end{equation}
\noindent where
$\nj$ and $a_j$ are the floating fit parameters,
$\Delta M$ is the bin width, and, depending on the
range, $a_2$ and/or $a_3$ can be set to be zero.

  The normalization scheme represented in Eq.~(4) will
be referred to as the efficiency (E) method.
The important features of the efficiency
method are that one makes no assumptions about
the composition of the non-resonant background,
and that it requires an absolute efficiency measurement for
the selection applied as well as absolute luminosity.
An alternate normalization scheme, which will be referred to as
the ratio (R) method, can be employed to 
probe several systematic effects. In the ratio method (which was
used in Ref.~\cite{babar}), instead
of using luminosity measured with $\diel$, $\gamma\gamma$,
and $\dimu$ events, we use the number of non-resonant
$\diel\to\gamma\dimu$ events underneath the 
$\jpsi$ signal in the muon pair mass distribution. 
This alternate integrated luminosity is 
\begin{equation}
\lumi_{\rm R}={{\dndm}\over{\ezero\,\dsdm}}\, ,
\end{equation}
\noindent in which $\dndm$ is the number of non-resonant
$\gamma\dimu$ events per unit mass from QED alone, evaluated 
at the $\jpsi$ mass (which in terms of our fit is
by definition the parameter $\azero$ from Eq.~(7)),
$\ezero$ is the efficiency for non-resonant QED muon pair
events to pass the selections,
and $\dsdm$ is the cross 
section per unit mass predicted by non-resonant QED alone, without detector
effects or selections, at 
$M$=$\mj$, including final state
radiation and vacuum polarization effects.
The value for $\dsdm$ is obtained by 
fitting the non-resonant $\diel\to\gamma\dimu$ MC
mass distribution at the generator level
with the polynomial portion of Eq.~(7) only, and dividing by the effective
luminosity of the MC sample; we find 
$\dsdm$=0.8510$\pm$0.0031~pb/MeV for the \BABAYAGA\ generator
and a value 1.7\% smaller when using a first-order
(at most, one photon per event) generator~\cite{BKJ}.
We compute $\ezero$ as in the first and third steps of that
for~$\epsilon$, but with \BABAYAGA\ FSR enabled, finding $\ezero$=69.2\%.

  The ratio method has, in place of Eq.~(4),
\begin{equation}
\left( {{\nj-\nb}\over{\azero}}\right)\left( {\ezero\over\epsilon}\dsdm\right)
 = \gamee\times\brmm\times \Iz\ .
\end{equation}
The ratio $\ezero/\epsilon$ is expected to be close to unity;
systematic effects which mostly cancel in this ratio
include those from trigger, reconstruction, radiative corrections,
and event selection variable modeling.
The ratio method replaces 
systematics associated with the standard
luminosity measurement and absolute efficiency determination 
with those applicable to $\azero$ and $\dsdm$.
We note that $\nj$ and $\azero$ are almost completely anti-correlated,
for which we account in the uncertainty propagation.

  Unlike the efficiency method, the ratio method requires
understanding non-$\jpsi$ backgrounds, \ie~the
extent to which other final states besides radiative
muon pairs populate the non-resonant entries in the mass distribution.
Based upon measured cross sections~\cite{LOWMUL} and MC studies 
of low-multiplicity
hadronic final states, we conclude
that all other backgrounds are negligible and assign 
an uncertainty of 0.3\% in $\brmm\times\gamee$ from this source.

\begin{figure}[thp]
\includegraphics*[width=5.5in]{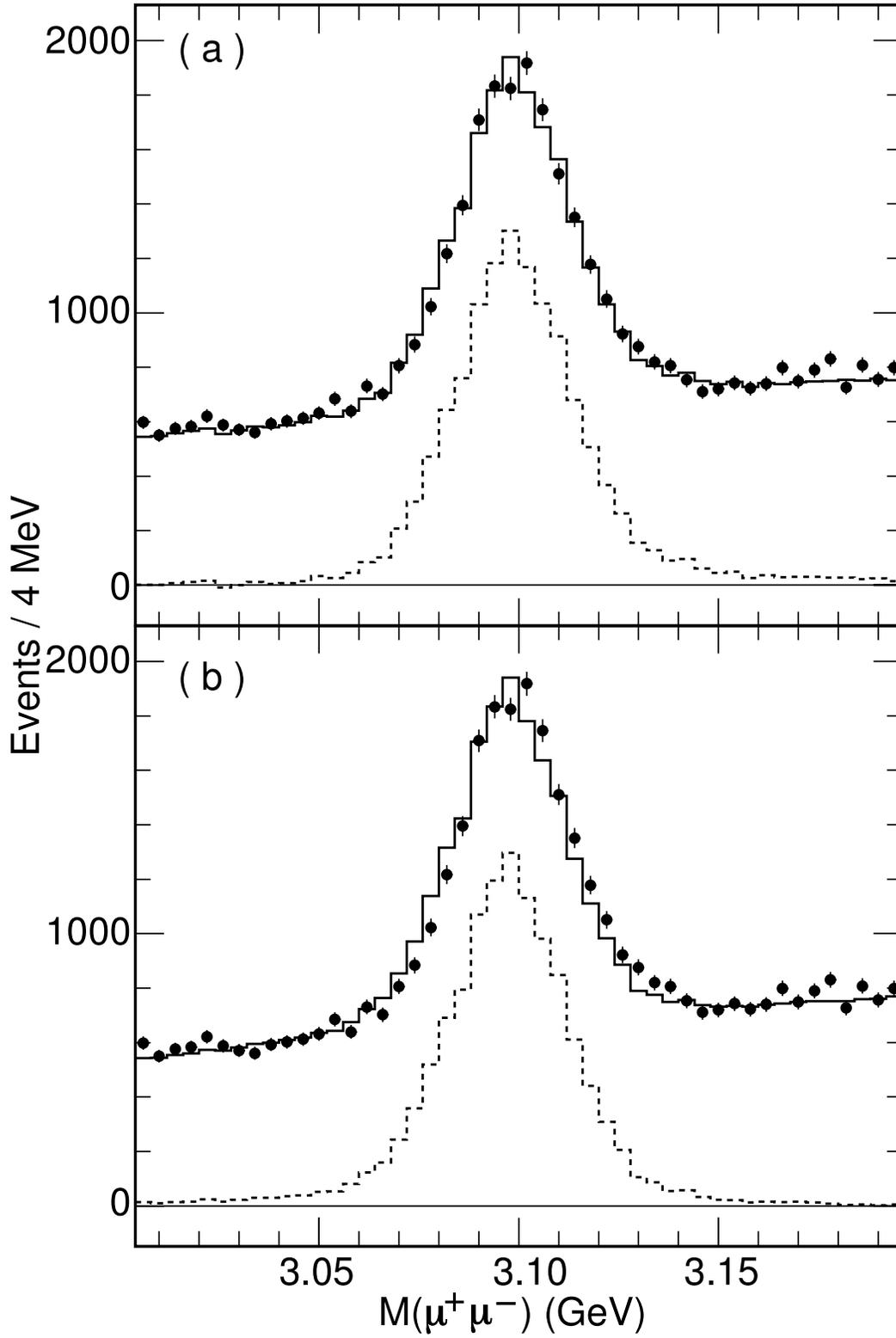}
\caption{Fit (solid line) of the muon pair invariant mass data (filled circles) 
to the sum of the
expected shape (dashed line) for a $\jpsi$ decay (a) with
and (b) without interference combined with
a smooth background (second order polynomial).\label{fig:intfit} }
\end{figure}

Fits over all mass ranges strongly prefer the shape that includes
interference to those which do not. 
This preference for the 2.9-3.3~GeV fit range,
as shown with interference in Fig.~\ref{fig:intfit}(a) 
and without in Fig.~\ref{fig:intfit}(b), 
is a 6.6$\sigma$ effect, as
determined from the difference in log-likelihoods from
the respective fits. The no-interference fit
also systematically underestimates the yield by
$\sim$3\%, which, not coincidentally, is 
the amount by which interference
changes the overall rate from 2.8-3.4~GeV.

  Table~\ref{tab:tableFits} lists the quantities
relevant to the $\brmm$$\times$$\gamee$ measurements.
Central values and statistical uncertainties of $\nj$ and $\azero$
are taken as unweighted means over the seven fits previously described,
which have CL's ranging from 1-18\%.
Combining information from fits over different mass ranges
in this manner samples different relative weightings
of background and signal regions. The 
systematic errors on $\nj$ and $\azero$ are taken
as the rms spreads of the corresponding fit results.
The values from the efficiency and ratio methods
are consistent within their uncorrelated uncertainties.
The polynomial fits and shape agree very well with the 
luminosity-normalized expectation from the radiative muon pair MC.

  For the efficiency method, the ``loose'' (``tight'') 
selections induce changes in $\brmm\times\gamee$
from nominal of $+$0.5\% ($-$0.0\%); similarly,
the ratio method variation is $+$0.0\% ($+$0.3\%).
These small changes demonstrate a good understanding of 
efficiency and background
levels, which are reflected in their systematic errors
in Table~\ref{tab:tableFits}.

The systematic errors on 
efficiency for the dimuon pair arise by extrapolating errors from
the CLEO $\psiprime\to X\jpsi$ analysis~\cite{xjpsiprl}
as appropriate.  Quoted errors for $\Is$ and $\dsdm$
include, in addition to MC statistical uncertainties,
contributions of 0.5\% to account for accuracy of the underlying
formulae.
The systematic error on $\nj$ (or $\nj/\azero$) from fitting,
computed as described above, is 0.9\% (or 1.8\%). 
The accuracy of the fitting assumptions is tested by pursuing the
fitting procedure using MC $\gamma\psiprime\to\gamma\dipion\dimu$
events for the fitting shape, MC $\gamma \jpsi$ events, without any interference,
for signal, and MC $\gamma\dimu$ for non-resonant background,
all with statistics much larger than those of the data.
From these high statistics samples, we find that the
above procedure introduces a bias in $\brmm\times\gamee$ of 1.0\% for the
efficiency method and 1.1\% for the ratio method, both in the
upward direction. We fully correct for these biases and assign
a 0.7\% systematic error to the corrections.
The relative uncertainty attributable to the statistics
of the resolution function in the data
was estimated as follows. An ensemble
$\gamma\psiprime\to\gamma\dipion\dimu$
MC samples was formed, each of the same
size as the corresponding sample from the data, and
each was used as the resolution function for 
the high-statistics MC muon pair distribution. The
relative rms variation of the fit results for $\nj$ ($a_0$) for the ensemble
was found to be 0.7\% (0.8\%), which is included
as an additional systematic uncertainty.
If, instead of adding the fitting systematic errors described above, 
we scale up the statistical errors by the square root
of the reduced-$\chi^2$ on each of the seven fits,
similar total uncertainties are obtained, indicating
reasonable error assignments.

  Reference~\cite{babar} asserts that $\ezero/\epsilon\equiv 1$ and
that the 
ratio method is insensitive to radiative corrections; we find
neither to be the case for our analysis due to FSR effects.
The ratio $\dsdm/\Iz$ changes by $\sim$5\%
when ignoring radiative corrections: when both ISR and FSR are allowed,
non-resonant events move from higher muon pair mass to lower mass,
resulting in a net increase in events at $M=\mj$, whereas decay
radiation in $\jpsi\to\dimu$ can only shift events
out of the signal peak to lower masses.

  We compute an error-weighted average of the E and R methods,
accounting for correlations. While the
ratio method avoids some of the systematics
of the efficiency method, it suffers a larger fitting error, because $\nj$ 
is almost fully anti-correlated with $\azero$
(which is the non-resonant muon pair level at $\mj$). 
This average gives relative weights of 8:1 for E:R. 
The weighted average is
$\brmm$$\times$$\gamee$=0.3384$\pm$0.0058$\pm$0.0071~keV.
For $\brmm$ we use the CLEO measurement~\cite{dilepprd}
$\brll$=(5.953$\pm$0.056$\pm$0.042)\%,
yielding $\gamee$=5.68$\pm$0.11$\pm$0.13~keV, and
$\gamtot$=95.5$\pm2.4\pm2.4$~keV.
In all cases the first errors quoted are statistical
and the second systematic, and the distinction 
between the two has been preserved in the
propagation of uncertainties in $\brll$ to $\gamee$ and $\gamtot$.

 This measurement of $\brmm\times\gamee$ is
consistent with the \BABAR~\cite{babar} value,
and the values determined here for $\gamee$ and $\gamtot$
are more precise and somewhat larger than all previous
measurements. 

  In summary, we have used the radiative return process
$\diel$$\to$$\gamma \jpsi$ to measure $\brmm$$\times$$\gamee$
with a 2.7\% relative uncertainty, 
and combined this with a CLEO measurement of $\brll$ to
obtain $\gamee$ (3.0\%) and $\gamtot$ (3.6\%) with
improved precisions. Combining with 
$\Gamma_{ee}[\psiprime]=2.54$$\pm$0.03$\pm$0.11~keV from Ref.~\cite{xj3770prl}
and accounting for common uncertainties of luminosity, 
$\brll$, and lepton tracking, 
we determine $\gamee[\psiprime]/\gamee(\jpsi)$=0.45$\pm$0.01$\pm$0.02 (5.0\%),
a quantity which might be more precisely predictable 
in LQCD than either $\gamee$ alone.

We gratefully acknowledge the effort of the CESR staff 
in providing us with excellent luminosity and running conditions.
A.~Ryd thanks the A.P.~Sloan Foundation.
This work was supported by the National Science Foundation
and the U.S. Department of Energy.

\begin{table}[thp]
\setlength{\tabcolsep}{0.4pc}
\catcode`?=\active \def?{\kern\digitwidth}
\caption{Number of background events expected.
}
\label{tab:tableBgd}
\begin{center}
\begin{tabular}{l|r}
\hline
Source & \# Events \\\hline
$\gamma \jpsi$, $\jpsi\to\dipion$  & 17.2 \\
$\gamma\psiprime\to\gamma\dipion \jpsi$, $\jpsi\to\dimu$         & 6.2 \\
$\gamma\psiprime\to\gamma\pi^0\pi^0 \jpsi$, $\jpsi\to\dimu$         & 62.4 \\
$\gamma\psiprime\to\gamma\eta \jpsi$, $\jpsi\to\dimu$               & 2.1 \\
$\gamma\psiprime\to\gamma\pi^0 \jpsi$, $\jpsi\to\dimu$              & 12.0 \\
$\gamma\psiprime\to\gamma\chi_{c0}$, $\chi_{c0}\to\gamma \jpsi$, $\jpsi\to\dimu$  & 0.5 \\
$\gamma\psiprime\to\gamma\chi_{c1}$, $\chi_{c1}\to\gamma \jpsi$, $\jpsi\to\dimu$  & 18.6 \\
$\gamma\psiprime\to\gamma\chi_{c2}$, $\chi_{c2}\to\gamma \jpsi$, $\jpsi\to\dimu$  & 49.6  \\
\hline
Total                       &168.7\\\hline\hline
\end{tabular}
\end{center}
\end{table}

\begin{table}[thp]
\setlength{\tabcolsep}{0.4pc}
\catcode`?=\active \def?{\kern\digitwidth}
\caption{Intermediate and final results for
the efficiency (E) and ratio (R) methods, with
statistical and systematic errors.
}
\label{tab:tableFits}
\begin{center}
\begin{tabular}{l|c}
\hline
Quantity & Value \\\hline
$\nj$                   & $12742\pm202\pm143$ \\
$\nb$                   & $169\pm9$          \\
$\Iz$~(pb/keV)          & $185.8\pm1.6$       \\
Efficiency (E) method & \\
\ \ \ \ $\epsilon$ (\%)         & $70.5\pm1.2$  \\
\ \ \ \ ${\cal L}$~(pb$^{-1}$)  & $280.7\pm2.8$       \\
\ \ \ \ Bias Factor             & $0.990\pm0.007$     \\
\ \ \ \ $\brmm\times\gamee$~(keV) &  
               $0.3385\pm0.0054\pm0.0075$     \\
Ratio (R) method & \\
\ \ \ \ $\azero$~(MeV$^{-1}$)      & $165.7\pm1.6\pm2.1$ \\
\ \ \ \ $\ezero/\epsilon$       & $0.981\pm0.008$     \\
\ \ \ \ $\dsdm$~(pb/MeV)        & $0.8510\pm0.0052$   \\
\ \ \ \ ${\cal L}_{\rm R}$~(pb$^{-1}$)& $281.5\pm5.3$       \\
\ \ \ \ Bias Factor             & $0.989\pm0.007$     \\
\ \ \ \ $\brmm\times\gamee$~(keV) &  
               $0.3373\pm0.0087\pm0.0096$     \\\hline
\hline
\end{tabular}
\end{center}
\end{table}

\end{document}